# Reliable Identifications of AGN from the WISE, 2MASS and Rosat all-sky surveys

Short title: AGN from all-sky surveys

R. Edelson [1, 2] & M. Malkan [3]

## ABSTRACT

We have developed the "$S_{IX}$" statistic to identify bright, highly-likely Active Galactic Nucleus (AGN) candidates solely on the basis of WISE, 2MASS and Rosat all-sky survey data. This statistic was optimized with data from the preliminary WISE survey and the SDSS, and tested with Lick 3-m Kast spectroscopy. We find that sources with $S_{IX} < 0$ have a $\gtrsim 95\%$ likelihood of being an AGN (defined in this paper as a Seyfert 1, quasar or blazar). This statistic was then applied to the full WISE/2MASS/RASS dataset, including the final WISE data release, to yield the "W2R" sample of 4,316 sources with $S_{IX} < 0$. Only 2,209 of these sources are currently in the Veron-Cetty and Veron (VCV) catalog of spectroscopically confirmed AGN, indicating that the W2R sample contains nearly 2,000 new, relatively bright (J $\lesssim$ 16) AGN.

We utilize the W2R sample to quantify biases and incompleteness in the VCV catalog. We find it is highly complete for bright (J < 14), northern AGN, but the completeness drops below 50% for fainter, southern samples and for sources near the Galactic plane. This approach also led to the spectroscopic identification of 10 new AGN in the Kepler field, more than doubling the number of AGN being monitored by Kepler. This has identified ~1 bright AGN every 10 square degrees, permitting construction of AGN samples in any sufficiently large region of sky.

*Subject headings:* Astronomical databases: surveys -- Galaxies: active -- Galaxies: Seyfert -- Methods: statistical -- Quasars: general

## 1. Introduction

Surveys for active galactic nuclei[4] (AGN), or extragalactic objects in general, have at least one of two broad goals: first, they can constrain the space densities and redshift-evolution of sources, providing key inputs to cosmological models. This requires a high level of completeness, or at least clear understanding of bias and incompleteness.

Second, such surveys often produce new samples of AGN useful for follow-up studies. For example, early AGN surveys (e.g., the 3C radio survey [Edge et al. 1959] and the ultraviolet-excess/spectroscopy surveys of Markarian et al. 1989 and Green et al. 1986) identified bright AGN that have subsequently been observed in many ways and at many wavelengths, providing key constraints on AGN and black hole physics. (Of course, those surveys have also made important contributions in their own right, on questions such as, e.g., AGN evolution.) Some AGN turn out to be of special value because of their location in the sky (e.g., near the ecliptic poles, and thus in the continuous viewing zone of some satellites) or due to their angular proximity to foreground galaxies (for absorption line studies), or to bright stars (for adaptive optics studies). For all of these purposes, completeness is not as important as the ability to confidently identify bright AGN throughout the sky.

---

[1] Department of Astronomy, University of Maryland, College Park, MD 20742-2421
[2] e-mail: rickedelson@gmail.com; phone: 970-531-3748
[3] Department of Physics and Astronomy, University of California Los Angeles, Los Angeles, CA 90095-1547

[4] In this paper, the term AGN refers only to objects with broad permitted emission lines (Seyfert 1-1.9 galaxies and quasars) or highly polarized objects (BL Lac objects and highly-polarized quasars), and not to narrow-line objects such as Seyfert 2 galaxies, LINERs and starburst galaxies.

In particular it would be useful to be able to construct a sample of AGN in any sufficiently large region of the sky, even if it has not been subjected to any spectroscopic surveys. That capability has until now been lacking. For example, the Kepler mission (Boruki et al. 2010) can produce unprecedented optical AGN light curves (e.g., Mushotzky et al. 2011), but it wasn't until after ~1.5 years of the mission had elapsed that a significant number of AGN in the Kepler field had been identified and monitoring begun.

This paper presents a method to identify AGN candidates solely on the basis of data from the Wide-field Infrared Survey Explorer (WISE; see Wright et al. 2010), the 2-Micron All-Sky Survey (2MASS; Skrutzkie et al. 2006) and the Rosat All-Sky Survey (RASS; Voges et al. 1999). It is arranged as follows: the next section presents the derivation of the $S_{IX}$ statistic to identify highly-likely AGN candidates from the preliminary WISE/2MASS/RASS data, and its initial optimization with Sloan Digital Sky Survey release 8 data (SDSS-8; York et al. 2000). Section 3 reports spectroscopic observations of new AGN candidates used to test the statistic, resulting in a catalog of highly-likely (≳95% confidence) AGN candidates across the sky. It also analyzes its effectiveness in terms of the intrinsic infrared and X-ray properties of AGN. Section 4 applies this statistic to the full WISE/2MASS/RASS data to derive the "W2R" catalog, and utilizes this sample to address a number of questions including the relative frequency of blazars, the completeness of the Veron-Cetty & Veron (2006; VCV hereafter) AGN catalog, and identification of new AGN in the Kepler field. Section 5 gives a brief concluding discussion. An Appendix presents an alternate AGN statistic that does not require detection in RASS, and the associated "W2" sample designed to maximize AGN completeness instead of reliability.

## 2. Method

### 2.1. Overview

AGN have distinctive continuum emission in both the infrared and X-ray bands. Few other sources emit strongly all the way from 1.2 to 22 μm (the full range of the WISE/2MASS data), and those that do have different colors from AGN. That is, it has been known for several decades that AGN have near- to mid-IR spectral energy distributions (SEDs) resembling a power law with a typical slope of $F_\nu \propto \nu^{-1}$ (Edelson & Malkan 1986). On the other hand, normal galaxies – stars plus cool dust – show ``bumpier'' SEDs dominated by starlight shortwards of 5 μm, and by cooler dust continuum and PAH bands from ~7 to 20 μm.

Likewise it appears that almost all AGN can be detected in the X-rays with sufficiently long integrations. For instance, ~80% of the 12 μm Seyfert 1s were detected in the RASS (Rush et al. 1996) even though the effective integration times were short (typically a few hundred seconds), and the remainder were detected in longer (~10 ks) pointed Rosat observations (Rush & Malkan 1996). These characteristics – distinctive infrared colors and X-ray detectability – make it possible to identify bright AGN solely on the basis of WISE, 2MASS and RASS data.

In this analysis, we are testing the null hypothesis that an object is not an AGN. All statistical tests such as this involve a tradeoff between Type I errors (incorrectly rejecting the null hypothesis when true, i.e. a 'false positive' inclusion of an object which is not a true AGN), and Type II errors (incorrectly accepting the null hypothesis when false; i.e., a 'false negative', where a true AGN is incorrectly excluded; see Fisher, 1935 for details).  Because we are primarily interested in high reliability, our approach minimizes the Type I error rate (contamination from inclusion of non-AGN), at the expense of a large (and ill-defined) Type II error rate (failure to include other real AGN).  That is, we are willing to tolerate a high degree of incompleteness in order to maximize our confidence in our AGN identifications.  (A complementary approach that minimizes Type II errors at the expense of a higher Type I error rate is given in Appendix A.)  Our statistic was empirically constructed by combining infrared colors and distance to the nearest X-ray source.  A low value of the resulting $S_{IX}$ index indicates an object with AGN-like 2–20 μm continuum and likely X-ray emission, corresponding to a high likelihood of being an AGN.

## 2.2. Defining the samples

To construct and test the AGN statistic, we begin by defining three test samples.  First, we restricted ourselves to the 83,510 objects detected at 10σ or better in all four WISE (3.4, 4.6, 12, 22 μm) and all three 2MASS (1.2, 1.6, 2.2 μm) bands.  This sample can be obtained by issuing the SQL command "w1mpro > 9 and w1snr > 10 and w2snr > 10 and w3snr > 10 and w4snr > 10 and n_2mass = 1 and j_msig_2mass < 0.1 and h_msig_2mass < 0.1 and k_msig_2mass < 0.1" to the WISE Preliminary Release Source Catalog on the page http://irsa.ipac.caltech.edu/cgi-bin/Gator/nph-scan.  Note that the first part of the command removes sources with W1 < 9 magnitudes.  As discussed in Section 4.1, this eliminates ~0.5% of the brightest AGN but nearly half of the stars, because very few AGN are brighter than W1 = 9.

WISE/2MASS associations are taken from the WISE preliminary database; in order to be included the separation must be less than 3" (see http://wise2.ipac.caltech.edu/docs/release/prelim/expsup/sec4_7.html#2mass_assoc for details).  For each object, we also measured the angular distance from the WISE position to the nearest RASS source.  We call this the "full" sample because it contains data on all currently available sources with good detections with both WISE and 2MASS.

In order to optimize the test statistic, we require a second sample of objects that have optical identification spectra.  We utilize a subsample of the 4,954 WISE/2MASS/RASS sources that also have SDSS-8 spectra.  We refer to this as the "SDSS" sample.  It is not a purely random sampling of the WISE/2MASS/RASS candidates, because SDSS was most focused on observing (often unusual) galaxies.  Relative to the full WISE sample, stars and "normal" spiral galaxies are underrepresented in the SDSS sample, while X-ray-bright galaxies (often AGN) are overrepresented.

The testing and validation also require a third subsample, containing just the sources identified as AGN in the SDSS sample.  To produce this, we visually examined each of the 797 spectra with SDSS class=QSO, isolating a sample of 741 with Seyfert 1/quasar spectra.  We refer to the 741 true broad-line active galaxies as the "lines=B" sample because these sources all exhibit broad emission lines.  The remaining 46 class=QSO sources turned out to be narrow-line objects, e.g., Seyfert 2s.  They had Hα emission lines mistakenly identified as "broad" due to poor automated continuum fitting, e.g. around the atmospheric absorption in the B band.  More generally, the other 4,157 SDSS sources were a combination of galaxies and stars.

## 2.3. Constructing the infrared color discriminator

The next step is to utilize these samples to construct a statistic that picks out AGN on the basis of their infrared colors.  The first three rows of Figure 1 compare histograms of the five measured infrared colors (J-H, H-K, W1-W2, W2-W3, W3-W4) for each of these three samples. (Following Wright et al. 2010, WISE bands centered at 3.4, 4.6, 11.6 and 22 μm are denoted W1, W2, W3 and W4, respectively.) The bottom row gives *R*, the ratio of the number of lines=B objects to the total number of SDSS objects in each color bin. For a given color bin, a value of *R* = 1 would mean that all of the sources in that bin are broad-line AGN, while *R* = 0 would indicate that all are non-AGN. Thus, *R* is a differential likelihood indicator that a source with a given color is an AGN.

The colors H-K, W1-W2, and W2-W3 show the largest range in *R*. (The underlying physical reason for this, discussed in Section 2.1., is that AGN have power-law infrared SEDs, while the infrared SEDs of normal galaxies are made up of multiple components, and the power-law is not that dissimilar from starlight in J-H and cool dust in W3-W4.) This is also apparent in Table 1: the AGN and non-AGN show very similar mean J-H and W3-W4 colors, while they are very different in the other three colors. Because of this, we formed our AGN template on the basis of the three colors H-K, W1-W2 and W2-W3, and ignored J-H and W3-W4.

Figure 1 reveals further important details of these color discriminators. Note that for W2-W3, the ratio *R* shows a fairly narrow and well-defined peak centered near ~2.93. We define a parameter to score how "AGN-like" that color is for each source, as follows:

$$c_4 = \text{abs}(2.93-(W2-W3)),$$

where abs refers to the absolute value function. A small value of c4 indicates a W2-W3 color close to the "optimal" value of 2.93, and a large value indicates a highly-discrepant value. Thus the smaller the score, the more likely this source is to be an AGN.

The parameter *R* behaves differently for H-K and W1-W2: in these cases, *R* increases to redder colors, with relatively broad transitions starting around ~0.77 and ~0.91, respectively. Thus for these two colors, we use a different functional form for the scoring parameter:

$$c_2 = \max(0.77-(H-K), 0),$$

$$c_3 = \max(0.91-(W1-W2), 0),$$

where max refers to the maximum function. That is, for H-K (W1-W2), any color redder than 0.77 (0.91) yields c2 (c3) = 0, while colors bluer than these receive increasingly positive scores, I e., higher penalties. Note that by design, the sense of all three "penalty functions" is the same: a low score indicates an AGN-like color and a high score indicates a non-AGN-like color, and all are positive-definite.

The observed AGN-like colors in each of these wavebands (Column 8 of Table 1) are very close to those that would have been expected from previous infrared photometry of well-known Seyfert 1 galaxies. Since the WISE and 2MASS magnitudes are on the Vega system, H-K = 0.77, W1-W2 = 0.91 and W2-W3 = 2.69 correspond to power-law slopes of approximately -1.07 from 1.66 μm to 2.16 μm, -0.80 between 3.4 and 4.6 μm, and -0.85 from 4.6 to 11 μm (see Table 1 in Wright et al. 2010). The near- to mid-infrared continua of Seyfert 1 nuclei are typically described by a $\nu^{-1}$ power-law. However, there are generally additional components present which alter the broadband colors, particularly starlight from the host galaxy which peaks at 1.6 μm, and a broad 3-5 μm ``bump" which makes the observed 3-12 μm slope significantly flatter (c.f. Edelson and Malkan 1986). The threshold colors we adopt encompass the mean AGN continuum shapes, as well as much of the cosmic scatter from one AGN to another.

Other groups have also used the approximate power-law shape of the AGN mid-IR continuum to select them, for example using 3.6—8um photometry from Spitzer/IRAC (Lacy et al. 2006, Stern et al. 2005). The idea that a red W1-W2 color measured by WISE should be a powerful selector for AGN was explicitly anticipated by Assef et al. (2010).

### 2.4. Optimizing the infrared color discriminator

Next, these scores are combined to produce a statistic with the same sense: a low score indicates a high likelihood that the source is an AGN, and the result is positive-definite. This is done by forming a linear combination of these three parameters and adjusting the weights to maximize the fraction of lines=B sources in the cohort of sources with the lowest scores. Because of the relatively small size of the lines=B sample, continuous optimization algorithms such as Newton's minimization method could not be used. Instead, this optimization was done iteratively by hand. This yielded the following figure of merit parameterization (only 2 of the 3 weights are independent) of the infrared colors, denoted $S_I$:

$$S_I = 1.7\, c_2 + 2.0\, c_3 + c_4.$$

Sources with infrared colors close to the AGN template have a low score, while those with large differences are scored high. This is seen in the first column of Figure 2, which shows a modest separation: most sources with $S_I < 0.888$ were non-AGN while most with $S_I > 0.888$ tend to be AGN. The larger weights for the 'color cut-off' functions c2 and c3 are needed, since those colors span a shorter wavelength baseline than does c4.

### 2.5. Constructing the RASS distance discriminator

Next we construct a second parameter $S_X$, based on the WISE-RASS source distance:

$$S_X = (d/e)^2,$$

where $d$ is the distance from the WISE source to the nearest RASS source and $e$ is the uncertainty in $d$ (both in arcseconds). A typical RASS source positional uncertainty is e=15. The ratio is a good indicator of the likelihood that the WISE source has an associated RASS source. The logarithm of this parameter is shown in the second column of Figure 2, which again shows good separation: a large fraction of sources with $\log_{10}(S_X) < 0.969$ were non-AGN while most with $\log_{10}(S_X) > 0.969$ were AGN. The sense of this parameter is the same as before: a smaller value indicates a higher likelihood of an associated X-ray source (and therefore that the source is an AGN), and it is positive-definite.

### 2.6. Synthesizing the final statistic

Because these parameters were derived independently (one based on infrared color, and the other on the distance to the nearest X-ray source), combining them should provide even sharper classification. Again, we made a linear combination with relative weighting by hand, to maximize the fraction of lines=B objects. The logarithm is then taken to collapse the values to a more manageable range. This yielded a combined statistic:

$$S_{IX} = \log_{10}(S_I + 0.116\, S_X),$$

This is shown in the third column of Figure 2. Note the very strong separation in the bottom diagram: of the 329 SDSS sources with $S_{IX} < 0$, 309 (94%) had lines=B, indicating an initial Type I error rate of ~6%.

### 3. Spectroscopic testing

### 3.1. Construction of the bright object test sample

The sources SDSS chose to observe were not drawn randomly from the larger population, as can be seen by the differences in the first two rows of Figures 1 and 2. For instance, note that relative to the SDSS sample, the full sample shows an excess of sources with H-K > 0.8 and/or W1-W2 > 0.5. This means that more field sources have favorable H-K and/or W1-W2 colors than the SDSS subsample, so it is necessary to make an independent test of the Type I error rate.

In order to test the error rate, it is it necessary to examine a fair (random) sample of the full sample, rather than those that SDSS selected for spectroscopy. Accordingly, we identified a subsample of the brightest northern W2R sources in the winter sky. Table 2 lists the 77 sources with $S_{IX} < 0$, J < 14.0, R.A. between $21^h$ and $15^h$, and Dec. > $-20^o$, sorted so that sources with the lowest (most favorable) value of $S_{IX}$ are at the top. Of these, 70 (91%) were listed in VCV: 69 have VCV classifications consistent with our AGN definition: S1-S1.9 (57), BL (5), HP (4), Q (2), AG (1). One (UGC 7252) is classified as S2, this does not fit our limited definition of an AGN. Finally, 7 of the W2R sources meeting the above criteria were not listed in VCV.

### 3.2. Spectroscopy with the Lick Observatory 3-m Telescope

In order to complete identifications of this sample, we observed these seven remaining sources on December 19 and 20, 2011 with the Kast double spectrograph on the Shane 3-m reflector. The blue channel covered wavelengths from the atmospheric cutoff to the dichroic cutoff at 5500 Å; the red channel covered from 5600 to 8200 Å. The slit width was 3"; the spectrum was extracted over a length of 5" to 7". Typical exposure times were 600 to 1200 seconds. Reductions to one-dimensional spectra, including bias-subtraction, flat-fielding and sky subtraction, were done using standard procedures in the twodspec apall packages of IRAF. Flux calibration, made by comparison with standard stars observed during twilight, is generally accurate to about 20%.

These spectra are presented in Figure 3, again sorted by $S_{IX}$. We find that all seven are AGN: six Seyfert 1s/quasars and one BL Lac object (W2R 0814-10). We also confirmed the VCV identification of UGC 3752 as a Seyfert 2. With these data, the entire bright AGN test sample has been fully identified. This indicates that all but one of the 77 sources with $S_{IX} < 0$ are AGN, indicating a formal Type I error rate of 1/77=1.4%. Note, however, that because the error on unity is unity, and it would be more accurate to conservatively quote a Type I error rate of ≲5%, corresponding to a success rate of ≳95%.

### 3.3. Extrema of $S_X$ and $S_I$

A cutoff at $S_{IX} < 0$ implies that $\log_{10}(S_I + 0.116 S_X) < 0$ and thus $S_I + 0.116 S_X < 1$. Since both $S_I$ and $S_X$ are positive definite, this means that both $S_I$ and $0.116 S_X$, must each be less than 1. Each of these has a simple, convenient interpretation. First, since $S_X = (d/e)^2$, this implies $d < 0.116^{-1/2} e$, or $d < 2.9 e$. That is, the infrared source must have an X-ray source within ~3σ or it cannot be included in the W2R sample irrespective of its infrared colors.

Second, the limit $S_I < 1$ means that the source must have "AGN-like" infrared colors. As discussed above, AGN show a characteristic power-law continuum between 2 and 12μm. This AGN component fills in the "valley" in the SED that is otherwise left between the coolest red giants (whose emission peaks at 1.6 μm) and the warmest dust grains (which emit strongly in the PAH emission bands around 8−11 μm, due to non-thermal spiking in very small grains). The AGN infrared component has been described as thermal re-radiation of the continuum from the central engine by hot dust grains outside the BLR (e.g., Pott et al. 2010). However, it is also well-known that a nonthermal synchrotron component produces a very similar power-law spectral shape (e.g. Edelson & Malkan 1986). Regardless of what its physical origin may be, this power-law mid-infrared continuum appears to be present in virtually every galactic nucleus which also has clearly detectable broad emission lines. It is also seen in some galaxies which are classified as Seyfert 2s, or even LINERs, probably because they too have a broad-line region (BLR), which is obscured from direct view. Since the extinction is less than a tenth of that in the optical, the method should work even in heavily dust-obscured galactic nuclei, even where optical methods could fail. However these obscured AGN will tend not to be detected in the X-rays, so they do not generally make it into this sample.

Starburst galaxies (i.e., lacking a true nonstellar AGN) have some hot dust which may be detectable at 3 μm, or even 2 μm, but it produces relatively much weaker continuum than the cooler dust. Therefore, very few starbursts have the mid-infrared power-law continua characteristic of AGN. On the other hand, nearly all AGN have it, so long as the strength of the AGN, relative to the underlying starlight of the host galaxy, is high enough.

Although some blazars probably lack a BLR, and may also lack this "hot" AGN component in the near-to-mid IR, they nonetheless have another, non-thermal, infrared power-law, which is known (from its high polarization and violent variability) to be direct synchrotron emission from a relativistically-beamed jet. Thus it is a fortunate coincidence that our same IR-color method is effective at picking out special galactic nuclei with either Seyfert 1 or blazar components.

## 4. Applications of the $S_{IX}$ statistic to AGN Searches

### 4.1. Derivation of the W2R sample

The $S_{IX}$ statistic was optimized and tested with preliminary WISE data covering ~57% of the sky. The final WISE data release on 13 March 2012 made it possible to apply this technique to the full sky with superior data. Accordingly, we applied the SQL command "w1mpro >= 9 and w1snr >= 10 and w2snr >= 10 and w3snr >= 10 and w4snr >=10 and n_2mass = 1 and j_msig_2mass <= 0.10866 and h_msig_2mass <= 0.10866 and k_msig_2mass <= 0.10866" to the WISE All-Sky Catalog at http://irsa.ipac.caltech.edu/cgi-bin/Gator/nph-scan to yield a sample of 884,556 sources. Note that this formulation differs slightly from that in Section 2.1., but these small improvements should not make any significant difference.

The main difference between the preliminary and final WISE data releases is that the latter catalog determines fluxes from "second pass" rather than "first pass" PSF-fitting. For the typical (i.e., faint) sources near the limits of our SNR cuts, the second pass PSF fit gives more accurate fluxes (see Figure 7 in http://wise2.ipac.caltech.edu/docs/release/allsky/expsup/sec6_3c.html). In the W1 band, the new fluxes are typically about 10% lower. The resulting colors are hardly changed from those given in the Preliminary release, but the SNR=10 limit has now become slightly more restrictive than before.

This application resulted in a sample of 4,316 sources with $S_{IX} < 0$. Table 3 lists these sources, referred to as the W2R AGN sample, for WISE/2MASS/RASS. By comparison, the preliminary WISE database yielded only 1,924 sources with $S_{IX} < 0$, so use of the final data release produced an improvement of a factor of ~2.2 in sample size.

The SQL command restricts the WISE sample to objects fainter than W1 = 9 in order to screen out a large number of stars. This was necessary both to facilitate data management as well as to minimize contamination from stars when searching for AGN in the Galactic plane.

This criterion however also eliminates a few dozen very bright AGN. The brightest AGN in the W2R sample is NGC 5548, which has W1 = 9.16. For comparison NGC 4151, the brightest AGN in the sky at many wavelengths has W1 = 6.90. For a roughly Euclidean flux/count relation, this difference of 2.26 magnitudes corresponds to ~23 missed sources. Thus the absence of some famous, very bright AGN, e.g., 3C 273 and all of the original Seyfert (1943) galaxies except NGC 5548 should not be taken as an error in our procedures. As discussed in the conclusions, we plan to investigate relaxing this cutoff in future work.

### 4.2. "Completeness" of the VCV catalog

We used the VCV spectroscopic galaxy classification to organize these sources into four subclasses: Seyfert 1 AGN (VCV classifications "S," "S1-S1.9," "Q," and "AG"), blazar AGN (VCV classifications "BL" and "HP"), non-AGN, and unclassified (not in the VCV catalog). (By our restricted definition, VCV classification "S2," and "S3", which includes LINERs and even some starbursts, all count as non-AGN). Only 2,209 of the 4,316 sources in the full list (51%) were in VCV at the time of this experiment. However, of the 77 sources in AGN test sample (those with J < 14 in a region of the northern sky), 70 (91%) are classified as AGN in VCV. Thus even though the VCV catalog was never intended to be a complete catalog of AGN, it nonetheless provides a very good census of bright (J < 14) AGN. It is less than half as complete for sources ~2 magnitudes fainter. It also reflects the much greater completeness of AGN searches in the northern hemisphere due to more extensive spectroscopic capabilities of ground-based telescopes there over the last several decades.

Further, we note that the sample of 2,209 identified sources breaks down into 2,104 Seyfert 1s, 87 blazars, and 18 non-AGN. Thus Seyfert 1s make up 2,104/(2,104+87) ~ 96% of the identified AGN sample and blazars make up the other ~4%. The formal Type I error rate is again small, 18/2,104 ~ 1%. But this is also a highly uncertain estimate, this time because an unknown bias may cause a different fraction of the 49% of unidentified sources to be non-AGN.

In order to test the completeness of the VCV catalog to AGN, Figure 5 shows the relative sizes of each of these four subsamples as a function of source brightness, declination and galactic latitude. Note that the completeness of the VCV catalog decreases strongly to fainter J magnitudes. Completeness also decreases to lower (more southern) declinations, with a fairly strong transition near the celestial equator. There is a hint of an increase in incompleteness at very northern declinations as well, but this is only seen in one data point. Finally there is also a clear increase in incompleteness near the galactic plane (small values of absolute magnitude of galactic latitude, |b|).

### 4.3. New Kepler AGN

One early motivation for this study was to identify AGN that could be observed with Kepler, as that satellite is producing light curves with precision and sampling that are more than an order of magnitude better than the best obtainable from the ground (see Mushotzky et al. 2011 for details). That satellite requires sources to be previously identified before their data can be downloaded, but because the field lies near the galactic plane, only two bright AGN (Zw 229-15 and 1RXS J192949.7+462231, both Seyfert 1 galaxies) were known at the start of that mission. Our group utilized the methods of Stocke et al. (1983), Malkan (2004), and an earlier incarnation of the $S_{IX}$ technique to identify and begin to monitor an additional 13 AGN in the Kepler field.

These now have spectroscopic confirmations from the Kast double spectrograph on the Lick 3-m telescope. The observations, obtained in July 2011 were reduced in the same way as described above for the December 2011 run, except that the slit width was 2.0", affording slightly better spectral resolution. These spectra are shown in Figure 6, and source details are given in Table 4. Note that the first 10 of the confirmed AGN in Table 4 are members of the W2R sample, the next (W2 1925+50) fell slightly outside the $S_{IX} < 0$ cutoff, and the last two are not in either sample. Note also that W2R 1926+42 is a BL Lac object.

## 5. Conclusions

It has long been known that AGN have distinctive properties in both the X-rays and infrared. This paper combines available data from three all-sky surveys – WISE, 2MASS and RASS – in order to develop the "$S_{IX}$" statistic that can identify AGN with high ($\gtrsim$95%) confidence. The method was optimized with preliminary WISE data and SDSS-8 spectra and independently tested with the Lick 3-m Kast spectrograph. It was then applied to the final WISE data release to produce the "W2R" AGN sample contains nearly 2,000 new, bright (J $\lesssim$ 16) AGN.

While neither this sample nor the VCV catalog claims to be "complete," it is possible to use each to test and quantify the biases and incompleteness in the other. We find that the VCV catalog presents a good census (~90% completeness) for northern (declination > -20°), bright (J < 14.0) sources, but the completeness falls below 50% for southern, fainter sources.

The success of this approach also indicates that "classic" AGN – Seyfert 1s, quasars and blazars – share two common properties. First, they tend to be strong soft X-ray sources. Second, their infrared SEDs are strongly dominated by a smooth, relatively flat power-law. This distinguishes them from so-called "obscured AGN" – Seyfert 2s, LINERs, ULIRGs, etc. – that have steep and bumpy infrared SEDs and are generally not detected in the soft X-rays.

Finally, we hope to make further improvements on this technique in the future. For instance the $S_{IX}$ statistic was derived using WISE preliminary data but applied to the WISE final data release. We plan to re-derive it using the final dataset and then confirm the results with spectroscopy at Lick over the next year. We will investigate relaxing the W1 > 9 brightness cutoff in order to increase the number of very bright AGN. We will also utilize Lick spectroscopy to confirm new AGN candidates from the final survey in the Kepler field. Finally we will examine the much larger and less reliable W2 sample, given in the appendix below.

## Acknowledgements

The authors would like to thank Richard Mushotzky and Wayne Baumgartner for insights into using X-ray data to identify AGN, and Tom McGlynn for help with the HEASARC program Xamin. We also thank the anonymous referee and Joan Wrobel, the Astrophysical Journal Scientific Editor, for a timely and helpful review of this paper. This research has made extensive use of data obtained from the High Energy Astrophysics Science Archive Research Center (HEASARC), provided by NASA's Goddard Space Flight Center, the NASA/IPAC Infrared Science Archive (IRSA), operated by the Jet Propulsion Laboratory, California Institute of Technology, under contract with NASA, and the Sloan Digital Sky Survey (please see http://www.sdss.org/collaboration/credits.html for details).

## Appendix: The W2 (WISE/2MASS) sample

Here we modify of our analysis to derive a sample with a good Type II error rate instead of the main focus of the paper, which is to derive a sample with a good Type I error rate. That is, the method and sample presented in this section is focused on identifying almost all of the lines=B sources, at the expense of a relatively large number of false positives.

There are two main differences between the two statistics and samples. The first is that RASS data are not used in this analysis, because only 410 of the 741 SDSS lines=B sources (~56%) show a RASS source within 3σ of the infrared position. (While it appears that all or almost all AGN can be detected with sufficiently long exposures, the typical ~350 sec RASS integrations were just not long enough to detect all sources.) Thus only $S_I$ is used in this analysis. The second difference is that a more liberal cutoff was used because the goal is to minimize the Type II errors. We drew a cutoff at $S_I < 1.73$.

Of the 741 SDSS objects with lines=B, 701 (95%) have $S_I < 1.73$, corresponding to a Type II error rate of ~5%. There are 1,255 SDSS sources with $S_I < 1.73$, indicating a success rate of 701/1,255 = 56% (at best). Of the full sample of 884,556 sources, 55,050 (6.2%) have $S_I < 1.73$. This is referred to as the W2 (for WISE/2MASS) sample. Its value is that it comes closer than the W2R sample to producing a complete sample of the AGN population, albeit at the cost of a far larger, and ill-determined Type I error rate.

Of these 55,050 sources, 4,774 have VCV identifications as AGN. This yields a lower limit on the success rate of 4,774/55,050 = 9%. Thus the Type I error rate is currently rather poorly-constrained to the range 44-91%. We plan to obtain Lick 3-m spectroscopy in the coming year in order to directly determine the W2 Type I error rate. Finally note that 543 of the W2 sources have VCV identifications of S2/S3, suggesting that 542/(4,774+543) = 10% of the identified W2 sample are narrow-line objects. This is much larger than the 0.5% of narrow-line sources in the identified W2R sample, confirming that the presence of a nearby RASS source selects against narrow emission line galaxies.

**Table 1: Mean colors and parameters for the three input samples**

| Color/Parameter | Full | | | SDSS | | | lines=B | | | Contrast |
|---|---|---|---|---|---|---|---|---|---|---|
| | Mean | Median | S.D. | Mean | Median | S.D. | Mean | Median | S.D. | |
| J-H | 0.94 | 0.82 | 0.48 | 0.75 | 0.75 | 0.13 | 0.77 | 0.78 | 0.17 | 0.20 |
| H-K | 0.54 | 0.50 | 0.30 | 0.53 | 0.50 | 0.19 | 0.77 | 0.78 | 0.22 | 1.34 |
| W1-W2 | 0.37 | 0.28 | 0.37 | 0.43 | 0.32 | 0.32 | 0.91 | 0.94 | 0.25 | 2.17 |
| W2-W3 | 3.01 | 3.20 | 1.16 | 3.61 | 3.72 | 0.55 | 2.93 | 2.90 | 0.35 | 1.82 |
| W3-W4 | 2.34 | 2.28 | 0.71 | 2.24 | 2.24 | 0.39 | 2.33 | 2.32 | 0.25 | 0.25 |
| $S_I$ | 2.57 | 2.64 | 1.22 | 2.21 | 2.52 | 0.96 | 0.62 | 0.45 | 0.58 | 2.71 |
| $\log_{10}(S_X)$ | 3.30 | 3.44 | 0.96 | 2.74 | 3.10 | 1.10 | 1.30 | 0.51 | 1.75 | 1.82 |
| $S_{IX}$ | 2.41 | 2.51 | 0.83 | 1.91 | 2.17 | 0.86 | 0.83 | 0.17 | 1.32 | 1.84 |

**Table notes**: The first column gives the measured color (for the first five rows) or parameterization (for the last three). The next nine columns give the mean, median and standard deviation for each color (or parameter), for the full WISE/2MASS/RASS sample, the SDSS subsample, and the lines=B subsample, respectively. The last column gives the "contrast" between the SDSS and lines=B samples; that is, the absolute value of the difference between the samples' median color (or parameter) divided by the mean standard deviation. Note that the colors J-H and W3-W4 show a relatively small contrast, but the other three colors show a large contrast. This makes those colors, H-K, W1-W2 and W2-W3 better-suited for distinguishing AGN from other objects in the Sloan survey.

**Table 2: Bright AGN test sample**

| Row | Source Name | RA | Dec | J | $S_{IX}$ | $S_I$ | $\log_{10}(S_X)$ | Type | z | 2MASS ID | RASS (1RXS J...) |
|---|---|---|---|---|---|---|---|---|---|---|---|
| 1 | 3C 120 | 04 33 11.08 | +05 21 15.6 | 12.67 | -1.319 | 0.019 | -0.602 | S1.5 | 0.033 | 776090812 | 043311.2+052112 |
| 2 | MARK 9 | 07 36 57.01 | +58 46 13.5 | 13.08 | -1.212 | 0.045 | -0.852 | S1.5 | 0.039 | 1057769027 | 073657.0+584610 |
| 3 | IRAS F07144+441 | 07 18 00.60 | +44 05 27.2 | 13.32 | -1.129 | 0.072 | -1.688 | S1.5 | 0.061 | 585728155 | 071800.7+440527 |
| 4 | IRAS 04416+1215 | 04 44 28.77 | +12 21 11.7 | 13.93 | -1.097 | 0.067 | -0.954 | S1n | 0.089 | 1283847063 | 044428.6+122113 |
| 5 | NPM1G-05.0216 | 04 47 20.71 | -05 08 14.0 | 13.63 | -1.003 | 0.064 | -0.510 | S1.5 | 0.044 | 197517295 | 044720.4-050813 |
| 6 | MARK 1383 | 14 29 06.59 | +01 17 06.1 | 13.04 | -0.957 | 0.094 | -0.852 | S1.0 | 0.086 | 1259029691 | 142906.7+011708 |
| 7 | MARK 79 | 07 42 32.81 | +49 48 35.0 | 12.55 | -0.899 | 0.067 | -0.292 | S1.2 | 0.022 | 880738049 | 074232.9+494830 |
| 8 | NPM1G-14.0512 | 13 41 12.89 | -14 38 40.1 | 13.24 | -0.893 | 0.039 | -0.116 | S1n | 0.042 | 1253848597 | 134112.5-143836 |
| 9 | SNU J04255+3447 | 04 25 33.10 | +34 47 19.5 | 13.63 | -0.866 | 0.107 | -0.602 | S1 | 0.058 | 1311860102 | 042533.0+344715 |
| 10 | RXS J04520+4932 | 04 52 04.76 | +49 32 44.8 | 13.09 | -0.824 | 0.112 | -0.486 | S1 | 0.029 | 813145631 | 045205.0+493248 |
| 11 | MCG -02.14.009 | 05 16 21.21 | -10 33 41.3 | 13.19 | -0.745 | 0.151 | -0.602 | S1 | 0.028 | 115706031 | 051621.5-103341 |
| 12 | RXS J03370+4738 | 03 37 02.91 | +47 38 50.2 | 13.57 | -0.692 | 0.087 | 0.000 | S1 | 0.184 | 421944963 | 033703.9+473852 |
| 13 | UM 614 | 13 49 52.85 | +02 04 45.2 | 13.67 | -0.670 | 0.208 | -1.306 | S1.8 | 0.033 | 866975302 | 134952.7+020446 |
| 14 | PG 0844+349 | 08 47 42.44 | +34 45 04.6 | 13.41 | -0.658 | 0.213 | -1.203 | S1.0 | 0.064 | 190208819 | 084742.5+344506 |
| 15 | B2 0321+33 | 03 24 41.15 | +34 10 46.0 | 13.93 | -0.640 | 0.113 | 0.000 | S1n | 0.063 | 174522289 | 032441.3+341056 |
| 16 | PG 1448+273 | 14 51 08.78 | +27 09 26.9 | 13.64 | -0.622 | 0.123 | 0.000 | S1n | 0.065 | 694625518 | 145108.5+270933 |
| 17 | RXS J06021+2828 | 06 02 10.46 | +28 28 19.4 | 12.23 | -0.609 | 0.208 | -0.486 | S1 | 0.033 | 149712032 | 060210.7+282821 |
| **18** | **W2R 0502+22** | **05 02 58.22** | **+22 59 51.8** | **13.23** | **-0.600** | **0.228** | **-0.704** | **S1** | **0.057** | **144694350** | **050258.5+225949** |
| 19 | HS 0306+1051 | 03 08 56.70 | +11 03 15.4 | 13.62 | -0.584 | 0.103 | 0.134 | Q | 0.150 | 846330760 | 030855.8+110320 |
| 20 | HS 0710+3825 | 07 13 40.27 | +38 20 39.9 | 13.35 | -0.571 | 0.180 | -0.116 | S1n | 0.123 | 850552011 | 071339.7+382043 |
| 21 | RXS J04344+7127 | 04 34 29.11 | +71 28 02.0 | 13.80 | -0.551 | 0.236 | -0.408 | S1.5 | 0.025 | 909346139 | 043429.0+712757 |
| 22 | MARK 124 | 09 48 42.68 | +50 29 31.5 | 13.85 | -0.543 | 0.201 | -0.134 | S1n0 | 0.056 | 972705125 | 094841.6+502926 |
| 23 | NGC 985 | 02 34 37.81 | -08 47 15.9 | 12.74 | -0.543 | 0.265 | -0.736 | S1.5 | 0.043 | 103425681 | 023438.0-084714 |
| 24 | PKS 0422+00 | 04 24 46.83 | +00 36 06.3 | 12.90 | -0.532 | 0.254 | -0.468 | HP | | 773008941 | 042446.8+003559 |
| 25 | S5 0716+71 | 07 21 53.44 | +71 20 36.2 | 11.94 | -0.527 | 0.252 | -0.408 | BL | | 575064129 | 072153.2+712031 |
| 26 | SNU J03569+4255 | 03 56 57.01 | +42 55 40.5 | 13.54 | -0.434 | 0.326 | -0.444 | S1 | 0.066 | 814116054 | 035657.3+425536 |
| 27 | IRAS 03450+0055 | 03 47 40.18 | +01 05 14.0 | 12.99 | -0.421 | 0.019 | 0.492 | S1.5 | 0.031 | 778498712 | 034738.7+010543 |
| 28 | MARK 1044 | 02 30 05.51 | -08 59 53.1 | 12.46 | -0.393 | 0.397 | -1.203 | S1n | 0.017 | 103338235 | 023005.5-085951 |
| 29 | MARK 813 | 14 27 25.05 | +19 49 52.4 | 13.86 | -0.385 | 0.374 | -0.486 | S1.0 | 0.111 | 1099804429 | 142725.3+194954 |
| 30 | B3 0754+394 | 07 58 00.04 | +39 20 29.0 | 12.91 | -0.372 | 0.381 | -0.422 | S1.5 | 0.096 | 850696538 | 075759.7+392036 |
| 31 | S5 2116+81 | 21 14 01.14 | +82 04 48.2 | 13.83 | -0.365 | 0.410 | -0.736 | S1.0 | 0.086 | 902844140 | 211400.0+820447 |
| 32 | MCG -02.12.050 | 04 38 14.18 | -10 47 44.9 | 13.73 | -0.365 | 0.398 | -0.538 | S1.0 | 0.036 | 136924508 | 043813.8-104740 |
| 33 | MS 04124-0802 | 04 14 52.65 | -07 55 39.6 | 12.73 | -0.363 | 0.194 | 0.315 | S1.5 | 0.037 | 144094444 | 041451.8-075521 |
| 34 | IRAS 06269-0543 | 06 29 24.76 | -05 45 26.4 | 13.41 | -0.327 | 0.290 | 0.194 | S1n | 0.117 | 129481404 | 062925.1-054535 |
| 35 | HS 1046+8027 | 10 50 35.67 | +80 11 50.7 | 13.71 | -0.310 | 0.134 | 0.486 | S1 | 0.115 | 852452906 | 105037.1+801204 |

**Table notes**: Sources that meet the following criteria: detected at >10σ in all WISE and 2MASS bands, $S_{IX} < 0$, J < 14, Right Ascension between $21^h$ and $15^h$, and Declination > $-20^o$. Of the 77 sources in this list, 70 are also listed in the VCV catalog. The remaining 7 sources were observed with the Lick 3-m; they are shown in boldface. The source name (taken from VCV when possible, otherwise denoted by a positional name) is given in the first column, followed by Right Ascension and Declination (taken from the WISE catalog) in Columns 2 and 3. The source identifications and redshifts (taken from VCV when available, otherwise derived from the spectra presented in Figure 3) are given in Columns 4 and 5. Newly-identified sources are shown in boldface and sources that do not meet our AGN criteria are shown in italics. The measured values of J are given in Column 6, and the parameterizations $S_I$, $\log_{10}(S_X)$ and $S_{IX}$ are given in Columns 7-9. The 2MASS identifier and RASS source name are given in Columns 10 and 11. Table is sorted in ascending order in $S_{IX}$.

**Table 2: Bright AGN test sample (continued)**

| Row | Source Name | RA | Dec | J | $S_{IX}$ | $S_I$ | $\log_{10}(S_X)$ | Type | z | 2MASS ID | RASS (1RXS J...) |
|---|---|---|---|---|---|---|---|---|---|---|---|
| 36 | MARK 110 | 09 25 12.85 | +52 17 10.5 | 13.87 | -0.310 | 0.284 | 0.250 | S1n | 0.035 | 783924768 | 092512.3+521716 |
| 37 | MARK 618 | 04 36 22.28 | -10 22 33.9 | 13.20 | -0.308 | 0.340 | 0.116 | S1.0 | 0.035 | 136884757 | 043622.3-102226 |
| 38 | 1WGA J0310+405 | 03 10 06.23 | +40 56 54.3 | 13.51 | -0.297 | 0.470 | -0.526 | BL | 0.137 | 380812710 | 031006.2+405700 |
| 39 | PKS 0829+046 | 08 31 48.88 | +04 29 39.1 | 13.04 | -0.290 | 0.260 | 0.338 | HP | 0.180 | 804079605 | 083147.7+043005 |
| 40 | VII Zw 118 | 07 07 13.09 | +64 35 59.0 | 13.67 | -0.265 | 0.522 | -0.736 | S1.0 | 0.079 | 535071464 | 070713.5+643558 |
| 41 | UGC 3478 | 06 32 47.16 | +63 40 25.4 | 12.94 | -0.260 | 0.485 | -0.250 | S1n | 0.012 | 518331815 | 063248.0+634026 |
| 42 | SNU J05547+6620 | 05 54 47.24 | +66 20 43.9 | 13.67 | -0.249 | 0.397 | 0.158 | S1 | 0.186 | 543721984 | 055448.8+662036 |
| 43 | PKS 0405-12 | 04 07 48.42 | -12 11 36.6 | 13.76 | -0.244 | 0.524 | -0.408 | S1.2 | 0.574 | 729336554 | 040748.7-121133 |
| **44** | **W2R 0535+40** | **05 35 32.12** | **+40 11 15.8** | **13.86** | **-0.243** | **0.526** | **-0.408** | **S1** | **0.019** | **459530968** | **053532.6+401116** |
| 45 | Z 0214+083 | 02 17 17.11 | +08 37 04.0 | 13.83 | -0.216 | 0.410 | 0.233 | BL | 1.400 | 1133315111 | 021716.0+083707 |
| 46 | PG 0804+761 | 08 10 58.63 | +76 02 42.5 | 12.98 | -0.210 | 0.595 | -0.736 | S1.0 | 0.100 | 1295367017 | 081059.0+760245 |
| 47 | MG J0509+0541 | 05 09 25.96 | +05 41 35.3 | 12.71 | -0.206 | 0.431 | 0.218 | BL |  | 780403237 | 050927.0+054145 |
| 48 | RXS J06080+3058 | 06 08 00.80 | +30 58 41.5 | 14.00 | -0.205 | 0.487 | 0.070 | S1 | 0.073 | 874128285 | 060801.7+305847 |
| 49 | 2E 0507+1626 | 05 10 45.53 | +16 29 58.1 | 13.39 | -0.194 | 0.581 | -0.292 | S1.5 | 0.017 | 1217738633 | 051045.4+162953 |
| **50** | **W2R 0413+72** | **04 13 37.62** | **+72 06 52.5** | **13.67** | **-0.179** | **0.647** | **-0.878** | **S1** | **0.105** | **547701979** | **041337.5+720649** |
| 51 | HS 0624+6907 | 06 30 02.51 | +69 05 03.9 | 13.25 | -0.179 | 0.574 | -0.116 | Q | 0.370 | 763053340 | 063001.7+690458 |
| 52 | Ton 1015 | 09 10 37.04 | +33 29 24.6 | 13.98 | -0.176 | 0.638 | -0.602 | BL | 0.354 | 205603292 | 091037.2+332920 |
| 53 | MARK 205 | 12 21 44.10 | +75 18 38.3 | 13.60 | -0.167 | 0.444 | 0.310 | S1.0 | 0.070 | 583065747 | 122144.4+751848 |
| 54 | MARK 374 | 06 59 38.09 | +54 11 47.6 | 13.51 | -0.163 | 0.425 | 0.352 | S1.2 | 0.044 | 504498144 | 065938.5+541136 |
| 55 | MARK 141 | 10 19 12.55 | +63 58 02.7 | 13.62 | -0.163 | 0.670 | -0.852 | S1.2 | 0.042 | 911287970 | 101912.1+635802 |
| 56 | MARK 10 | 07 47 29.06 | +60 56 00.8 | 13.34 | -0.162 | 0.680 | -1.088 | S1.0 | 0.030 | 760619126 | 074729.4+605601 |
| **57** | **W2R 0406+11** | **04 06 22.05** | **+11 52 15.6** | **13.94** | **-0.158** | **0.531** | **0.149** | **S1** | **0.033** | **1310414811** | **040621.6+115233** |
| *58* | *UGC 3752* | *07 14 03.84* | *+35 16 45.4* | *12.44* | *-0.126* | *0.524* | *0.285* | *S2* | *0.016* | *1230840575* | *071404.6+351622* |
| 59 | RX J03140+2445 | 03 14 02.72 | +24 44 32.8 | 13.82 | -0.123 | 0.290 | 0.602 | AG | 0.054 | 123793901 | 031401.2+244504 |
| **60** | **W2R 0228+28** | **02 28 55.33** | **+28 09 04.5** | **13.95** | **-0.122** | **0.615** | **0.083** | **S1** | **0.036** | **350481074** | **022857.8+280904** |
| 61 | RXS J05524+5928 | 05 52 28.15 | +59 28 37.0 | 12.53 | -0.120 | 0.553 | 0.250 | S1 | 0.058 | 723962675 | 055229.5+592842 |
| 62 | MARK 595 | 02 41 34.86 | +07 11 13.8 | 12.98 | -0.107 | 0.730 | -0.352 | S1.5 | 0.028 | 1155145092 | 024135.2+071117 |
| 63 | MARK 684 | 14 31 04.79 | +28 17 14.1 | 13.22 | -0.096 | 0.795 | -1.203 | S1n | 0.046 | 586673408 | 143104.8+281716 |
| 64 | NGC 1019 | 02 38 27.40 | +01 54 27.8 | 13.67 | -0.092 | 0.804 | -1.306 | S1.5 | 0.024 | 1098807494 | 023827.5+015426 |
| **65** | **W2R 0157+47** | **01 57 10.94** | **+47 15 59.2** | **13.67** | **-0.088** | **0.765** | **-0.352** | **S1** | **0.047** | **459180062** | **015711.8+471607** |
| 66 | NPM1G-16.0109 | 02 56 02.62 | -16 29 15.4 | 13.65 | -0.081 | 0.796 | -0.538 | S1.9 | 0.032 | 40931691 | 025601.7-162919 |
| 67 | IRAS 04312+4008 | 04 34 41.53 | +40 14 21.5 | 12.97 | -0.058 | 0.759 | 0.000 | S1n | 0.020 | 369140072 | 043442.1+401425 |
| 68 | 3C 66A | 02 22 39.60 | +43 02 07.8 | 12.64 | -0.053 | 0.844 | -0.457 | HP |  | 1316752374 | 022239.1+430220 |
| 69 | 1ES 0206+522 | 02 09 37.40 | +52 26 39.5 | 13.97 | -0.041 | 0.758 | 0.116 | S1 | 0.049 | 476576303 | 020936.8+522645 |
| **70** | **W2R 0814-10** | **08 14 11.69** | **-10 12 10.2** | **13.41** | **-0.039** | **0.819** | **-0.083** | **BL** |  | **213426127** | **081411.7-101200** |
| 71 | PKS 1424+240 | 14 27 00.40 | +23 48 00.1 | 12.68 | -0.038 | 0.900 | -0.852 | HP |  | 1099812234 | 142700.5+234803 |
| 72 | NPM1G+78.0048 | 12 42 36.13 | +78 07 20.4 | 13.75 | -0.035 | 0.511 | 0.550 | S1.9 | 0.022 | 987868938 | 124250.9+780738 |
| 73 | UGC 3901 | 07 33 53.11 | +49 17 31.0 | 13.98 | -0.027 | 0.883 | -0.310 | S1.9 | 0.022 | 972501261 | 073353.4+491737 |
| 74 | IRAS 04576+0912 | 05 00 20.76 | +09 16 55.7 | 13.85 | -0.025 | 0.517 | 0.565 | S1n | 0.037 | 968100516 | 050022.3+091659 |
| 75 | UGC 6728 | 11 45 15.95 | +79 40 53.4 | 13.10 | -0.025 | 0.941 | -1.688 | S1.2 | 0.006 | 846701200 | 114516.1+794054 |
| 76 | NPM1G-03.0212 | 04 46 56.74 | -03 05 11.7 | 13.62 | -0.021 | 0.952 | -1.803 | S1 | 0.031 | 197501982 | 044656.8-030513 |
| 77 | RXS J03000+1630 | 03 00 07.99 | +16 30 14.5 | 13.76 | -0.002 | 0.708 | 0.393 | S1 | 0.035 | 800837315 | 030007.6+163023 |

Table 3: W2R sample

| Row | 2MASS ID | RA | Dec | J | $S_{IX}$ | $S_I$ | $\log_{10}(S_X)$ | RASS Name | VCV Name | Type | z |
|---:|---:|---|---|---|---|---|---|---|---|---|---|
| 1 | 4356519 | 12 46 46.81 | -25 47 49.1 | 14.384 | -2.309 | 0.004 | -2.108 | 1RXS J124646.7-254749 | PKS 1244-255 | HP | 0.638 |
| 2 | 977486214 | 07 53 01.38 | +53 52 59.8 | 15.467 | -2.016 | 0.003 | -1.243 | 1RXS J075301.1+535305 | S4 0749+54 | BL | 0.2 |
| 3 | 145276761 | 07 38 07.02 | +21 27 29.6 | 15.343 | -1.986 | 0.000 | -1.051 | 1RXS J073806.9+212726 | | | |
| 4 | 249700394 | 12 25 27.18 | -04 18 57.1 | 14.738 | -1.913 | 0.008 | -1.438 | 1RXS J122527.1-041857 | RXS J12254-0418 | S1 | 0.137 |
| 5 | 762994872 | 03 05 36.50 | +76 22 56.2 | 15.268 | -1.891 | 0.008 | -1.380 | 1RXS J030536.7+762257 | | | |
| 6 | 740172497 | 07 03 24.24 | -60 15 22.5 | 13.294 | -1.885 | 0.011 | -1.757 | 1RXS J070324.1-601525 | | | |
| 7 | 617850898 | 22 48 41.16 | -51 09 53.1 | 14.565 | -1.841 | 0.000 | -0.906 | 1RXS J224841.4-510951 | 2RE J2248-510 | S1.5 | 0.102 |
| 8 | 169461995 | 02 38 45.14 | -30 48 23.3 | 13.530 | -1.838 | 0.005 | -1.086 | 1RXS J023845.4-304825 | ESO 416-G05 | S1.2 | 0.063 |
| 9 | 618425984 | 05 02 52.38 | -47 13 30.5 | 15.782 | -1.814 | 0.000 | -0.879 | 1RXS J050252.6-471326 | | | |
| 10 | 995435651 | 04 08 40.54 | +30 39 12.4 | 15.601 | -1.761 | 0.017 | -2.549 | 1RXS J040840.5+303912 | | | |

**Table notes**: Sources in the W2R sample of highly-likely AGN candidates. Table 3 is published in its entirety in the electronic edition of the Astrophysical Journal, and can also be downloaded from http://www.astro.ucla.edu/~malkan/w2r/table3.xls. A portion is shown here for guidance regarding its form and content. The 2MASS identifier is given in the first column, followed by Right Ascension and Declination in Columns 2 and 3. The measured values of J are given in Column 4, and the parameterizations $S_I$, $\log_{10}(S_X)$ and $S_{IX}$ are given in Columns 5-7. The RASS source name is given in Columns 8. If a source is listed in the VCV catalog, the source name, redshift and identification from that catalog are given in Columns 9-11. Previously unidentified are shown in boldface while those that do not fit this paper's definition of AGN are shown in italics.

**Table 4: Lick spectroscopy of Kepler field AGN**

| ID | RA | Dec | J | $S_{IX}$ | z | Kepler ID | Comments |
|---|---|---|---|---|---|---|---|
| W2R 1914+42 | 19 14 15.50 | +42 04 59.9 | 15.71 | -0.603 | 0.502 | 6595745 | Hβ,OIII |
| W2R 1920+38 | 19 20 47.75 | +38 26 41.3 | 16.04 | -0.382 | 0.368 | 3337670 | Hβ,OIII,MgII |
| W2R 1931+43 | 19 31 12.56 | +43 13 27.6 | 15.49 | -0.326 | 0.439 | 7610713 | Hα/N2 |
| W2R 1858+48 | 18 58 01.11 | +48 50 23.4 | 15.87 | -0.323 | 0.079 | 11178007 | Hα,Hβ,Hg |
| W2R 1904+37 | 19 04 58.65 | +37 55 41.0 | 14.34 | -0.309 | 0.089 | 2694185 | Hα/N2 |
| W2R 1910+38 | 19 10 02.50 | +38 00 09.6 | 15.76 | -0.275 | 0.130 | 2837332 | Hα/N2 |
| W2R 1853+40 | 18 53 19.28 | +40 53 36.4 | 15.24 | -0.219 | 0.625 | 5597763 | MgII,Hβ |
| W2R 1845+48 | 18 45 59.57 | +48 16 47.6 | 15.43 | -0.167 | 0.152 | 10841941 | Hα,NaD,HK |
| W2R 1931+38 | 19 31 15.49 | +38 28 17.2 | 16.02 | -0.136 | 0.158 | 3347632 | Hα/N2 |
| W2R 1926+42 | 19 26 31.05 | +42 09 59.0 | 15.59 | -0.076 | 0.154 | 6690887 | BL Lac  NaD,HK |
| W2 1925+50 | 19 25 02.18 | +50 43 13.8 | 14.23 | 0.218 | 0.067 | 12158940 | Hα/N2 |
| 1915+41 | 19 15 09.13 | +41 02 39.1 | 15.69 |  | 0.220 | 5781475 | Hα,OIII |
| 1922+45 | 19 22 11.23 | +45 38 06.2 | 15.92 |  | 0.115 | 9215110 | Seyfert 1.9 Hα/N2 |

**Table notes**: New AGN in the Kepler field. The last column gives the strongest spectral features identified in each spectrum, which were used to derive the redshift. In nearly all cases they are emission lines (except for the stellar absorption lines found in the one BL Lac object. The first 10 sources are in the W2R sample; they are sorted by S. The next fell slightly outside the cutoff but is in the W2 sample. The last two were not included in either sample; they were initially found by the techniques of Stocke et al. (1983) and Malkan (2004).

**Table 5: W2 sample**

| Row | 2MASS ID | RA | Dec. | J | $S_I$ | VCV Name | Type | z |
|---|---|---|---|---|---|---|---|---|
| 1 | 617850898 | 22 48 41.16 | -51 09 53.1 | 14.565 | 0.000 | 2RE J2248-510 | S1.5 | 0.102 |
| 2 | 520542968 | 19 20 17.61 | -47 11 13.5 | 14.671 | 0.000 | | | |
| 3 | 248255370 | 20 06 19.32 | +32 34 11.1 | 15.068 | 0.000 | | | |
| 4 | 145276761 | 07 38 07.02 | +21 27 29.6 | 15.343 | 0.000 | | | |
| 5 | 165172583 | 23 58 28.23 | -22 59 31.8 | 15.449 | 0.000 | | | |
| 6 | 217606410 | 10 04 16.57 | -07 46 34.6 | 15.581 | 0.000 | | | |
| 7 | 618425984 | 05 02 52.38 | -47 13 30.5 | 15.782 | 0.000 | | | |
| 8 | 1280655360 | 17 00 04.27 | -15 16 15.3 | 15.843 | 0.000 | | | |
| 9 | 179865237 | 05 28 36.55 | -15 19 24.8 | 15.895 | 0.000 | | | |
| 10 | 710146476 | 07 55 34.27 | -73 08 02.7 | 16.161 | 0.000 | | | |
| 11 | 396995135 | 00 10 24.64 | +51 52 34.2 | 13.718 | 0.001 | | | |
| 12 | 883197922 | 13 39 22.92 | -54 45 06.6 | 14.920 | 0.001 | | | |

**Table notes**: The "W2" sample of moderate-likelihood AGN candidates. Table 5 is published in its entirety in the electronic edition of the Astrophysical Journal, and can also be downloaded from http://www.astro.ucla.edu/~malkan/w2r/table5.xls. A portion is shown here for guidance regarding its form and content. The 2MASS identifier is given in the first column, followed by Right Ascension and Declination in Columns 2 and 3. The measured values of J are given in Column 4, and $S_I$ is given in Column 5. If a source is listed in the VCV catalog, the source name, redshift and identification from that catalog are given in Columns 6-8. The table is sorted primarily by increasing value of $S_I$ and secondarily by increasing J magnitude.

**FIGURE CAPTIONS**

Figure 1. The five columns give measured 2MASS/WISE colors (J-H, H-K, W1-W2, W2-W3, W3-W4). The first three rows are histograms showing the distribution of these colors for each the three samples considered in this paper (the full sample of 83,510 WISE/2MASS/RASS sources, the 4,954 with SDSS-8 data, and the 741 identified as lines=B). Each histogram has 50 bins. The last row gives $R$, the ratio of counts in the last two histograms. $R$ data are plotted for only bins with more than 5 counts in the SDSS histogram.

Figure 2. First four rows are the same format as Figure 1, for the parameterizations $S_I$, $\log_{10}(S_X)$ and $S_{IX}$ in columns 1-3. The fifth row is the integral probability, that is, the expected success rate for a sample of sources with a parameterization up to a given value. The vertical red dotted lines show where R crosses 0.5 (based on interpolation between the two nearest points) at $S_I = 0.888$, $\log_{10}(S_X) = 0.969$ and $S_{IX} = 0.316$. The cutoff used to define the W2R sample is shown as a green dashed line at $S_{IX} = 0$. Likewise the blue dashed line at $S_I = 1.73$ indicates the cutoff used to derive the W2 sample in Appendix A.

Figure 3. KAST red side spectra (5460-8200 A) of the 7 bright W2R sources not listed in VCV. The spectral response of the detector was removed by dividing the spectra by a flatfield of the illuminated dome ceiling. Thus the flux scale is arbitrary, but does preserve relative slope differences between the spectra. The curvature in some of the continua around 5600 Å is an artifact of inadequate correction for the cutoff of the D55 dichroic. The dotted blue vertical line shows the redshifted wavelength of the strongest expected absorption line, from Na D. The horizontal red dashed lines show the expected locations of the strongest emission lines, Hα, [NII]6584, the [SII]6716/6730 doublet, and [OIII]5007. The telluric absorption in the A and B bands at 6870 and 7650 Å is indicated in green. The Seyfert 1 galaxies are identified in all cases by the broad wings on their strong Hα emission line. The first six sources are clearly broad-line AGN. The last source, W2R 0814-10 has a featureless spectrum; its identification as a BL Lac object is based on a radio source (NVSS J081411-101208) within 2".

Figure 4. Plot of $\log_{10}(S_X)$ vs. $S_I$ for the SDSS sample. Lines=B AGN are shown as red crosses, other (non-AGN) sources are shown as blue x's. The solid black line in the lower left indicates the $S_{IX} = 0$ cutoff; the region below and left of that line has a ~6% contamination with non-AGN. For the RASS-detected objects (lower half of the graph), a less stringent choice of $S_I$ would accept more AGN, while gradually including more contaminants. However there are many AGN which are too X-ray weak for RASS to detect (upper left corner). These are also strongly concentrated to small values of $S_I$ (less than 1). But even adopting the strict cutoff $\log_{10}(S_I) < 0$ still includes a large minority of non-AGN contaminants.

Figure 5. Composition of the W2R sample as a function of J magnitude (top), declination (middle) and absolute value of Galactic latitude (|b|, bottom). The 4,316 sources were ranked and sorted into 26 bins of 166 objects each, and the median of each bin is given on the x-axis. Seyfert 1s/quasars are shown as blue circles, blazars are green x's, non-AGN as red crosses, and unidentified sources as black boxes. Note that the completeness is best for bright, northern sources away from the Galactic plane.

Figure 6. Kast optical spectra of the 13 new AGN in the Kepler field. The left two panels show the blue-side and red-side spectra for 6 objects. The remaining 7 have only their red-side spectra shown, since these have the key broad Balmer line emission which is crucial to identifying the Seyfert 1 (or quasar) nucleus. On the blue side some of these AGN show the CaII HK doublet from stellar absorption. Three at the highest redshifts show broad MgII 2800A line emission. The red dashed lines indicate the redshifted wavelength for various emission lines and the blue dotted lines for absorption lines, as in Figure 3. The source W2R 1926+42 has a featureless spectrum; it is identified as a BL Lac based on a radio source (NVSS J192631+420958) within 2".

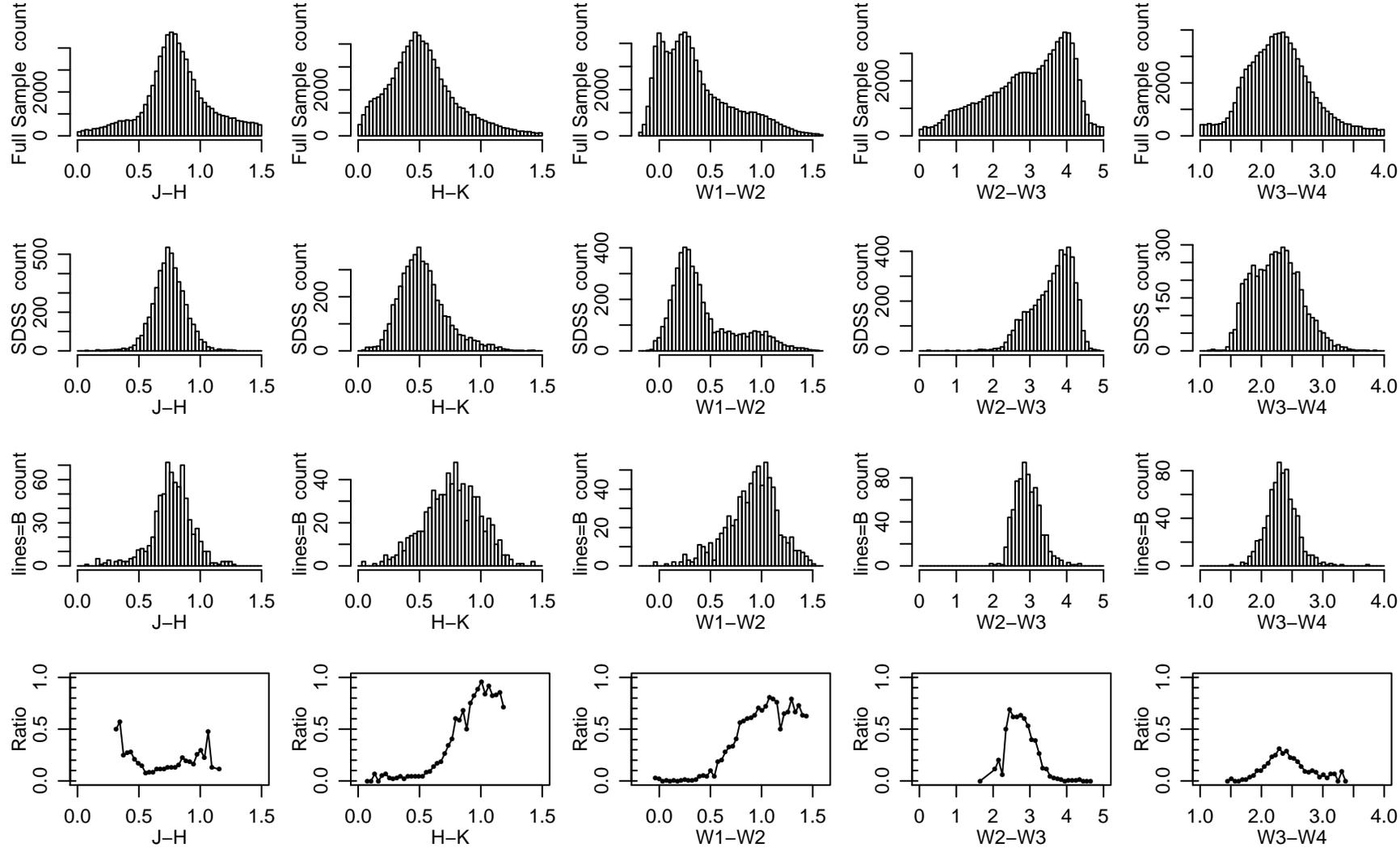

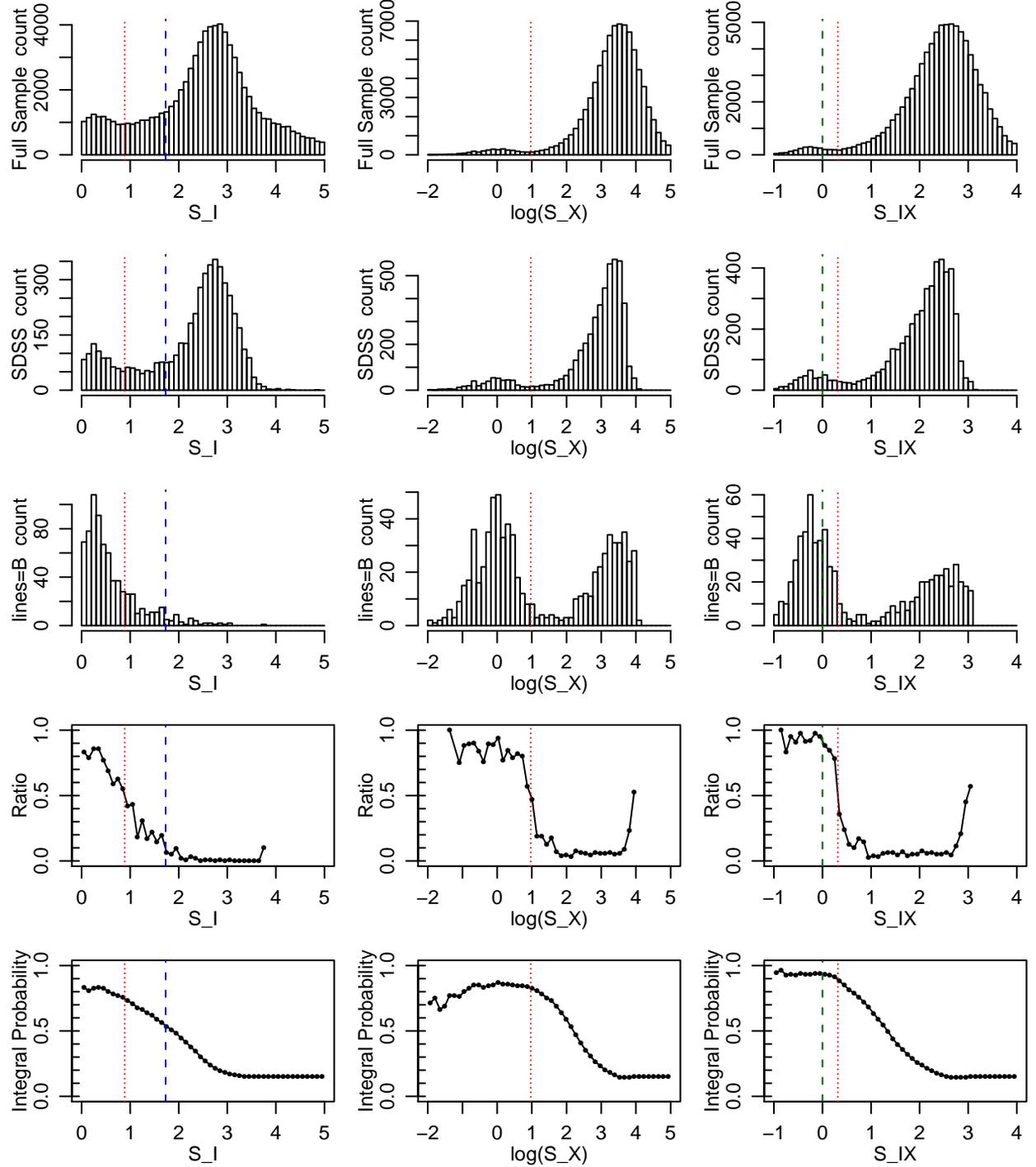

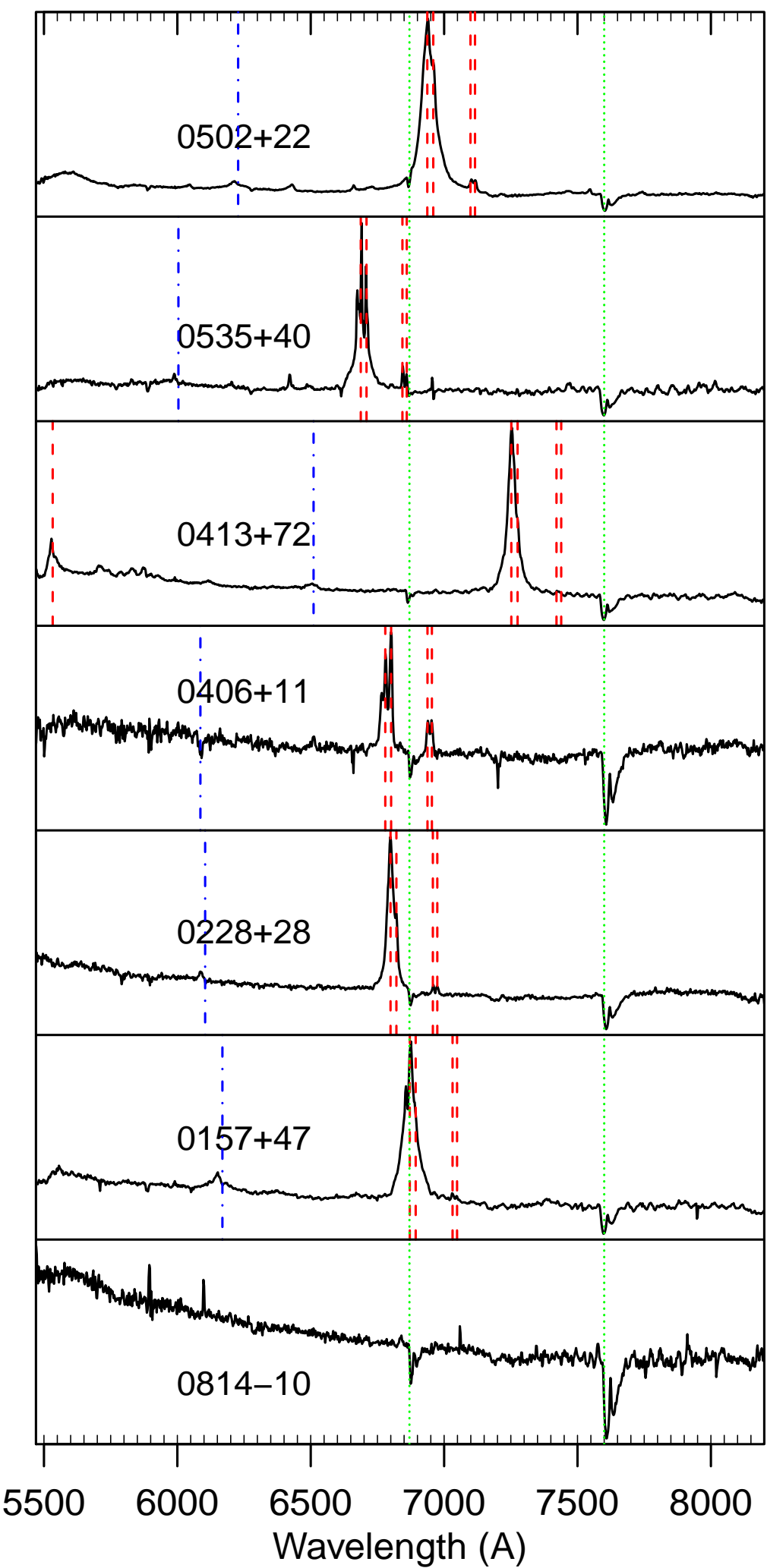

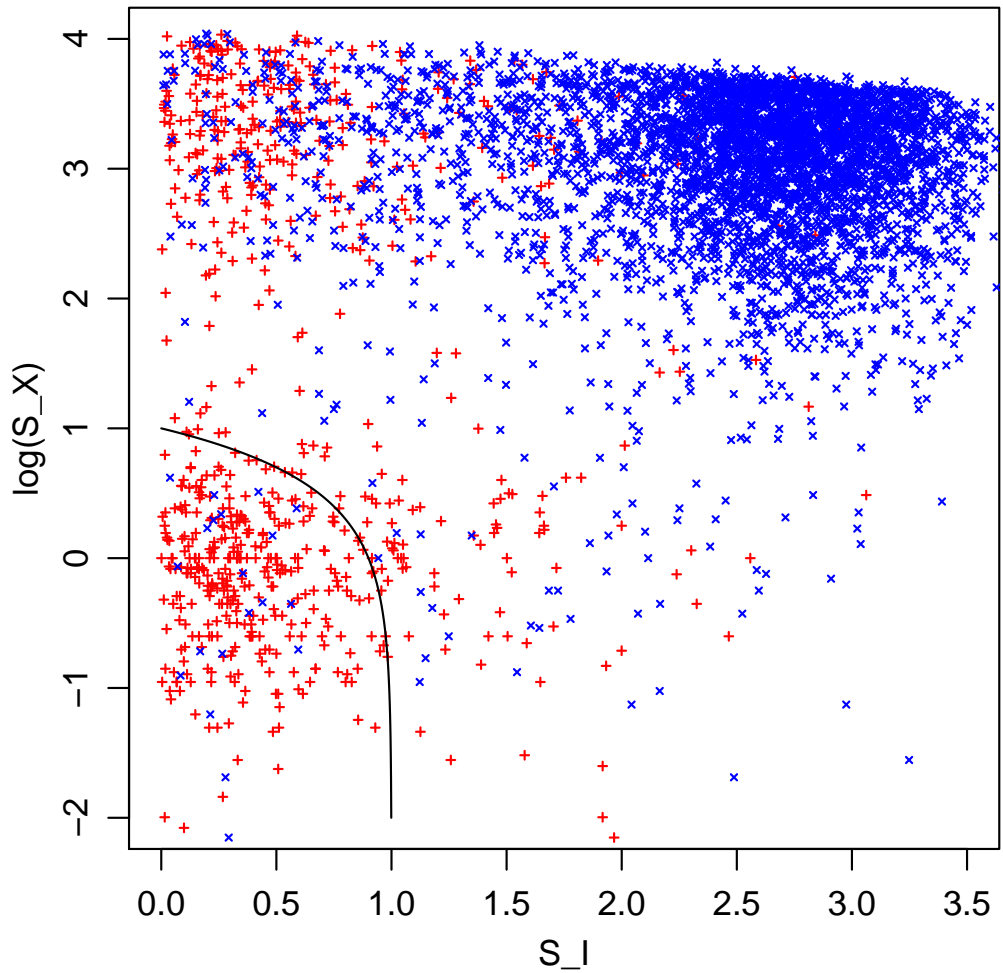

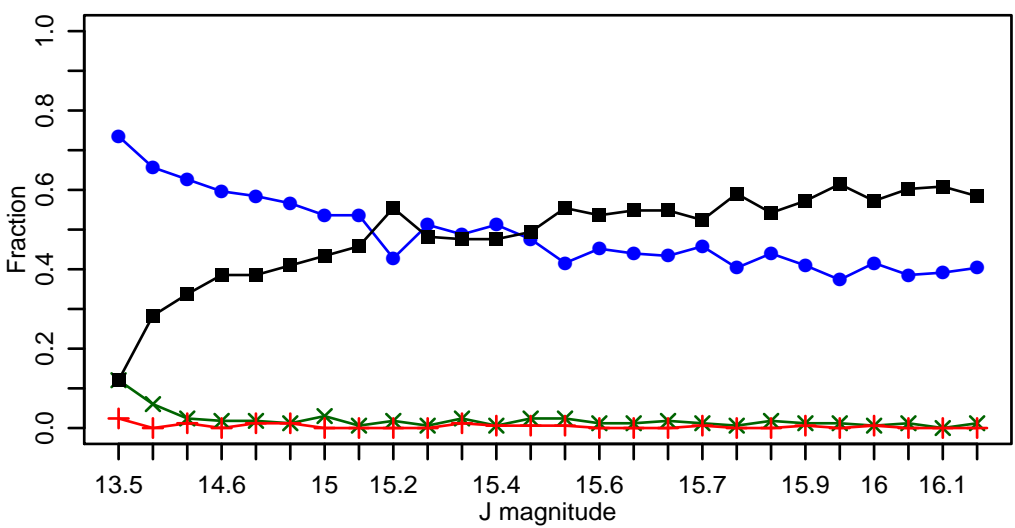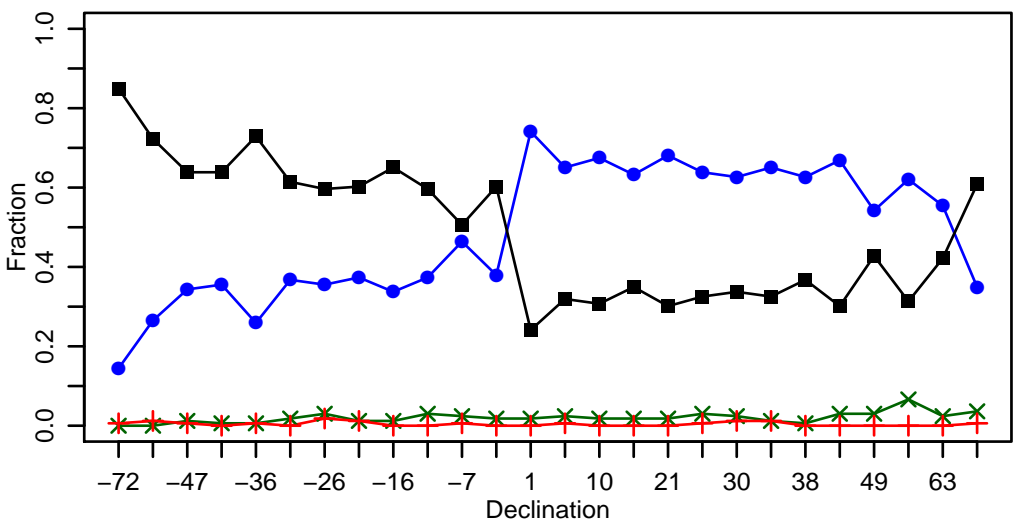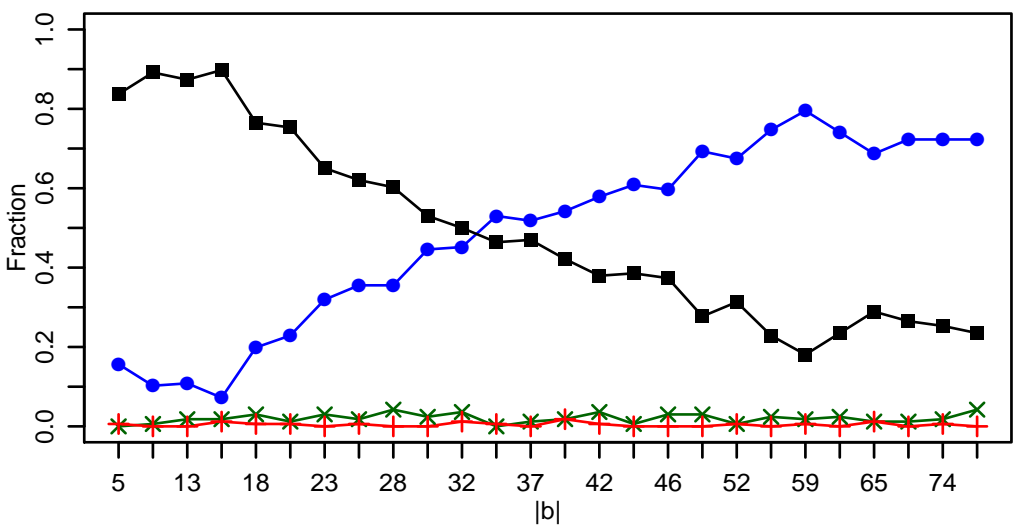

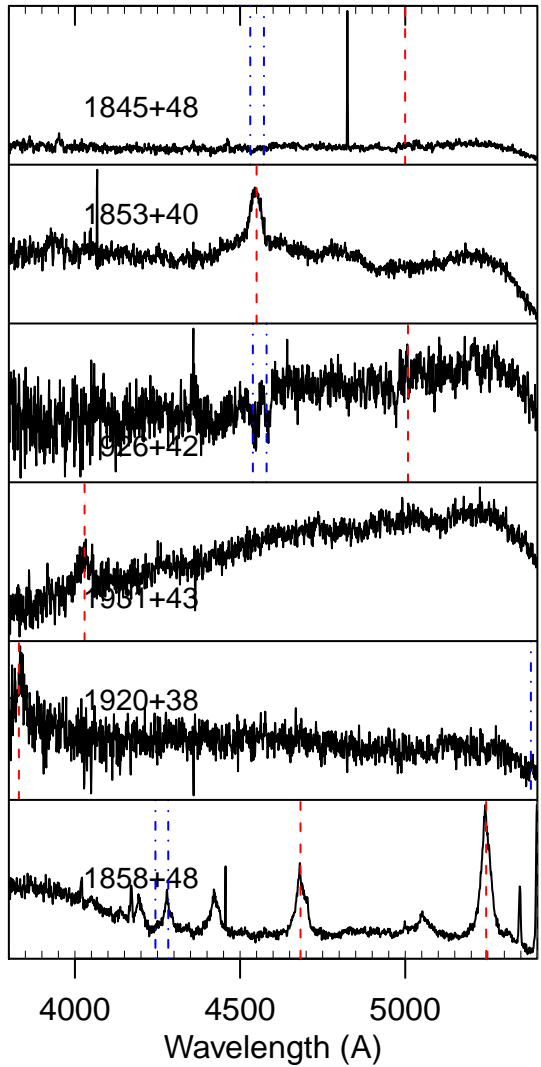
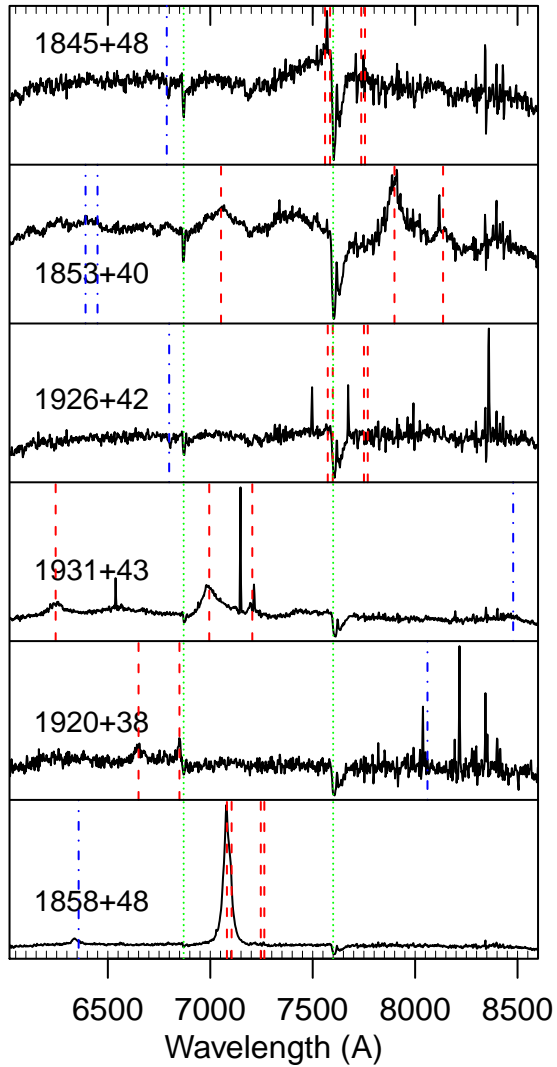
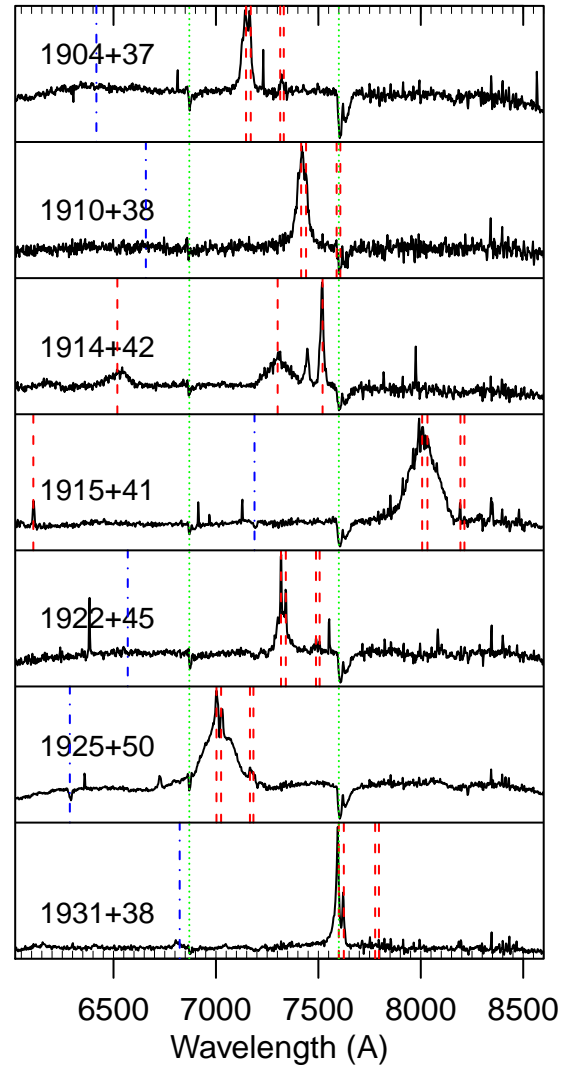